%% file: ms.tex
\documentclass[useAMS,usenatbib]{mn2e}
\usepackage{epsfig}
\usepackage{subfigure}
\usepackage{amsmath,amssymb}
\usepackage{float}
\usepackage{aas_macros}
\input{macro.tex}

\newcommand{\fig}{fig/}

\title{The velocity--shape alignment of clusters and the kinetic Sunyaev--Zeldovich effect}
\title[Velocity--shape alignment of clusters and the kSZ effect]
{The velocity--shape alignment of clusters and the kinetic Sunyaev--Zeldovich effect} 

\author[Cunnama et al.]{D. Cunnama$^{1,3}$\thanks{E-mail:cunnama@gmail.com}, A. Faltenbacher$^{2,4}$, C. Cress$^{1,3}$, S. Passmoor$^{1}$
\\
$^{1}$Physics Department, University of the Western Cape, Cape Town 7535, South Africa\\
$^{2}$Max-Planck-Institute for Astrophysics, Karl-Schwarzschild-Str. 1, D-85741 Garching, Germany \\
$^{3}$Centre for High Performance Computing, 15 Lower Hope St., Rosebank, Cape Town, South Africa\\
$^{4}$ MPA/SHAO  Joint  Center  for  Astrophysical  Cosmology  at
Shanghai Astronomical Observatory,\\ Nandan Road 80, Shanghai 200030, China}

\begin{document}

\date{\today}

\pagerange{\pageref{firstpage}--\pageref{lastpage}} \pubyear{0000}

\maketitle

\label{firstpage}

\begin{abstract}
We use the Millennium simulation to probe the correlation between cluster velocities and their shapes and the consequences for measurements of the kinetic Sunyaev-Zeldovich (kSZ) effect. Halos are generally prolate ellipsoids with orientations that are correlated with those of nearby halos. We measure the mean streaming velocities of halos along the lines that separate them, demonstrating that the peculiar velocities and the long axes of halos tend to be somewhat aligned, especially for the most massive halos.  Since the kSZ effect is proportional to the line-of-sight velocity and the optical depth of the cluster, the alignment results in a strong enhancement of the kSZ signature in clusters moving along the line of sight. This effect has not been taken into account in many analyses of kSZ signatures. 

\end{abstract}


\section{Introduction}
\label{sec:intro}
Observations of large-scale flows of matter in the universe provide strong constraints on structure formation theories and cosmology. In particular, observations of peculiar velocities of clusters of galaxies could constrain cosmological parameters \citep[e.g.,][ and references therein]{Bhattacharya-Kosowsky-08} or possibly challenge the $\Lambda$CDM paradigm \citep{Kashlinsky-08b}. 

The new generation of Cosmic Microwave Background (CMB) experiments such as the Atacama Cosmology Telescope (ACT, \citealt{Kosowsky-03}) and the South Pole Telescope (SPT, \citealt{Ruhl-04}) will identify thousands of galaxy clusters out to redshifts beyond $z\approx1$ via the thermal Sunyaev-Zeldovich (tSZ) effect \citep{Sunyaev-Zeldovich-72}. The tSZ effect in clusters arises from the inverse Compton scattering of CMB photons by free electrons in the intracluster medium (ICM) and involves a change of frequency of CMB photons, with radio photons typically being shifted to frequencies above 217 GHz . Another smaller effect, the kinetic Sunyaev-Zeldovich effect (kSZ), which is due to the peculiar velocity of electrons in the ICM produces a frequency independent distortion to the CMB spectrum. Attempts to measure the kSZ effect have mostly produced upper limits on the peculiar velocity of clusters (\citealt{Holzapfel-97,Benson-03}) but the current CMB experiments should detect the kSZ effect with high significance \citep[e.g.,][]{Hernandez-08a}.

Prospects for extracting peculiar velocities of clusters detected via the kSZ, and using these for cosmology, have been discussed by many authors: analytical work includes that by \cite{Rephaeli-Lahav-91}, \cite{Haehnelt-Tegmark-96}, \cite{Aghanim-01}, \cite{Holder-04}, \cite{Bhattacharya-Kosowsky-08}, \cite{Zhang-08} \cite{Hernandez-08a}; work based on full simulations include that by \cite{Yoshida-Sheth-Diaferio-01}, \cite{Diaferio-05}, \cite{Schafer-06a}, \cite{Schafer-06c} and \cite{Nagai-03}. The potential to cross-correlate kSZ signatures with other foreground tracers is discussed in \cite{Dore-04} and \cite{Dedeo-05}. We also note that, despite claims to the contrary \citep{Croft-04,Peel-06}, linear theory should describe cluster velocity correlations fairly well if the mean streaming motions of massive halos towards each other are included \citep{Sheth-Zehavi-08}. 

Almost all of the work on the extraction of kSZ signatures of clusters assumes spherical clusters but cluster halos are not perfectly spherical. In this paper we work out the correlation between the orientation of ellipsoidal galaxy clusters and their peculiar velocity in order to investigate the enhancement of the kSZ signal due to this correlation. The kSZ signal is proportional to the line-of-sight velocity and the optical depth of the cluster \citep{Sunyaev-Zeldovich-72}. On average an alignment between the orientation and velocity will result in a higher optical depth in clusters moving along the line of sight.

Section \ref{sec:sim} provides a brief description of the data utilized in our calculations. Section \ref{sec:shape} presents the correlation between the velocity and shape of the clusters and section~\ref{sec:ksz} describes the effect this correlation has on measurements of the kSZ signal. Section~\ref{sec:conclusion} provides some discussion and conclusions drawn from our results.
\section{Simulation and halo catalog}
\label{sec:sim}
Our halo catalog was extracted from the Millennium Simulation \citep{Springel-05a} which adopted roughly concordance values for the parameters of a flat $\Lambda$ cold dark matter ($\Lambda$CDM) cosmological model, $\Omega_{\rm dm}=  0.205$ and $\Omega_{\rm  b}= 0.045$ for the current densities in CDM and baryons, $h= 0.73$ for the present dimensionless value of the Hubble constant, $\sigma_8= 0.9$ for the rms linear mass fluctuation in a sphere of radius $8 \hMpc$ extrapolated to $z= 0$, and $n= 1$ for the slope of the primordial fluctuation spectrum.

The halos are found using a two-step procedure. The first step entails identifying all collapsed halos with at least 20 particles using a friends-of-friends (FoF) group-finder with linking parameter b = 0.2 times the mean particle separation.
The substructure algorithm  SUBFIND \citep{Springel-01} is then used to subdivide each FoF-halo into a set of self-bound {\it sub-halos}. In this study we exclusively focus on sub-halos which we simply refer to as {\it halos}.

We present results for halos within mass bins ranging from $10^{12}$ to $10^{13} \hMsol$, $10^{13}$ to $10^{14}\hMsol$ and $10^{14}$ to $10^{15}\hMsol$. There were 401334, 43139 and 2701 halos in each respective mass bin.
\section{Shape-velocity alignment}
\label{sec:shape}
\begin{figure*}
  \begin{minipage}[t]{0.45\hsize}
  \epsfig{file=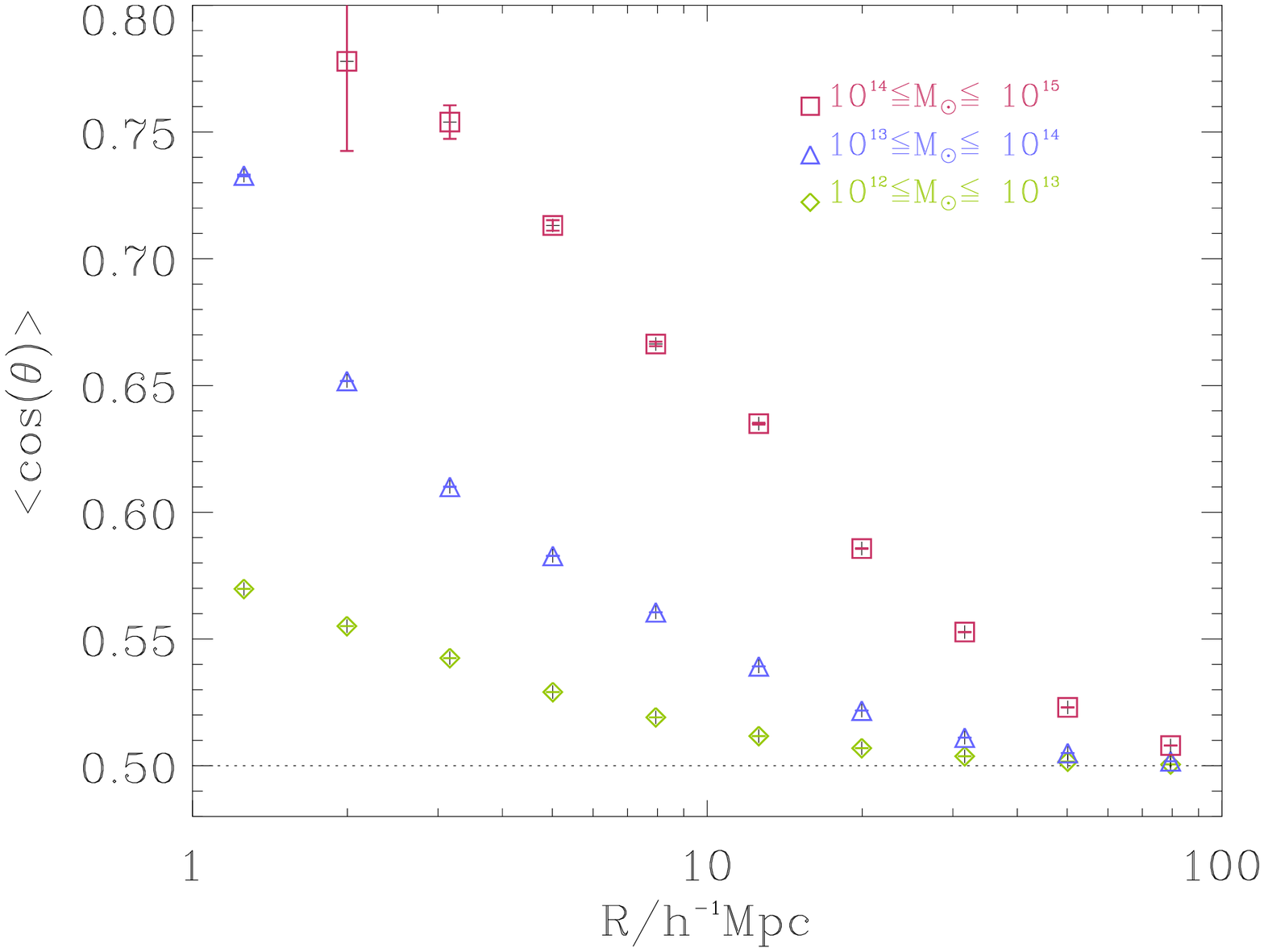,width=0.95\hsize }
   \caption{\label{fig:ori} Mean cosine  of angles, $\theta$, 
between halo  orientations and  the connecting line to neighbouring halos, for halos  within  the  indicated  mass  range, as a function of separation. The dotted line illustrates the results for an isotropic distribution. }
  \end{minipage}
  \hspace{0.7cm}
  \begin{minipage}[t]{0.45\hsize}
  \epsfig{file=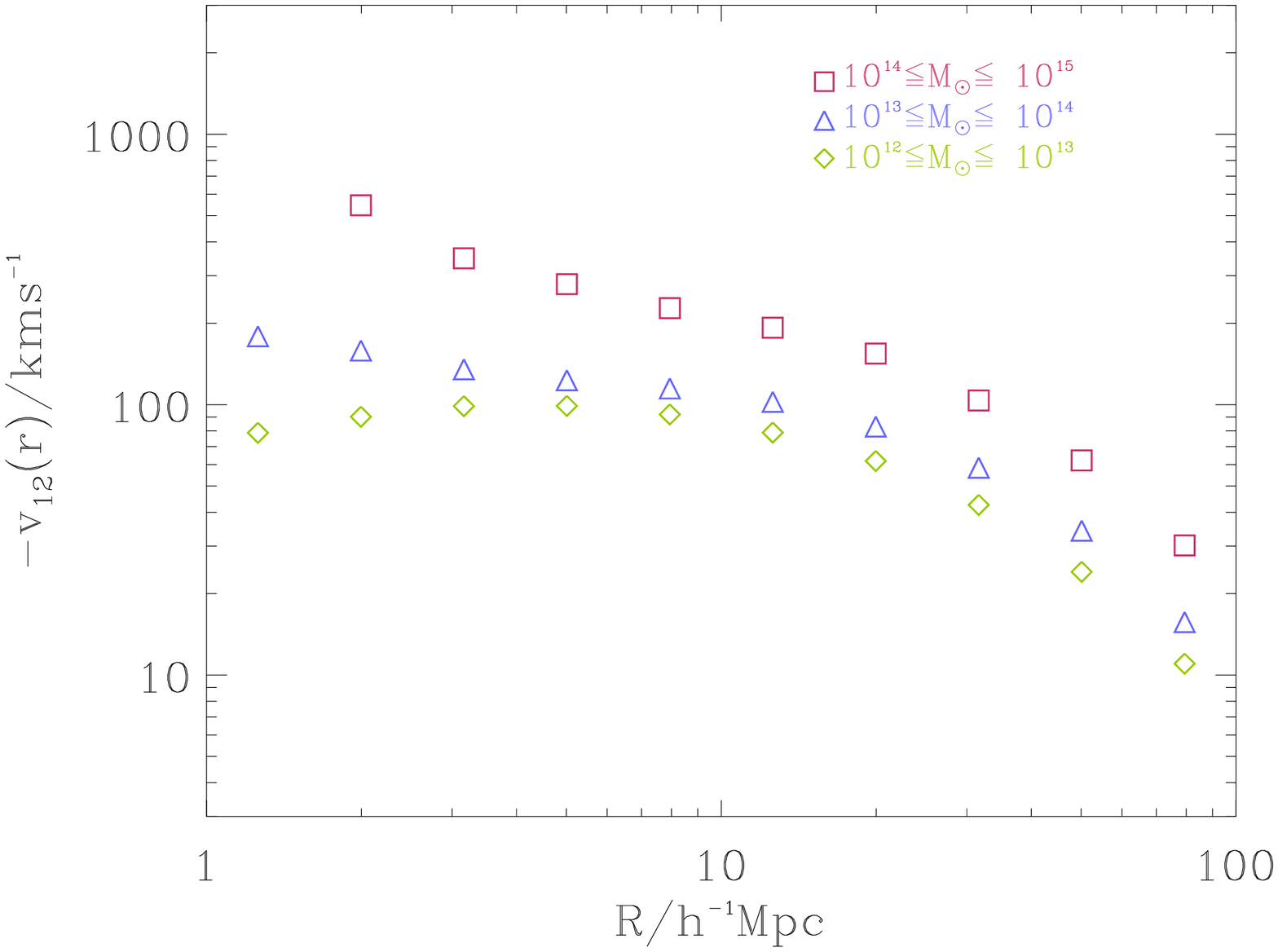,width=0.95\hsize }
  \caption{\label{fig:vel} Mean streaming velocities of halos along the line of separation within the indicated mass ranges.}
  \end{minipage}
\end{figure*}
Two general properties of clusters are well known. Firstly it has been shown that in general clusters tend to point towards each other \citep[e.g.,][]{Faltenbacher-02} and secondly that they tend to stream towards each other \citep{Sheth-Zehavi-08}. From these properties it can be anticipated that there should therefore be some correlation between the velocity of a cluster and its orientation.

To investigate such a correlation we extract these quantities from the halo catalog. The velocities of the halos are calculated by taking the peculiar bulk velocity of all particles belonging to the halo. As is standard, the orientations are derived by evaluating the eigenvalues of the inertial tensor, or more precisely, the second moment of the mass tensor of the halo. The maximum eigenvalue obtained corresponds to the major axis of the halo and the minimum to the minor axis.

To confirm the correlation between halo orientations and neighbouring halos we compute $\theta$, which is the angle between the major axis of a halo and the the connecting line to another. We then bin the angles derived for each pair according to $r$, which is the separation between the two halos. Finally we determine the mean cosine for logarithmic separation bins and calculate the mean angle for various separations of halos. The results are shown in Figure~\ref{fig:ori} for halos within the indicated mass range. A mean cosine $>0.5$ indicates alignment between connecting line and orientation. For separations of $\lesssim50\hMpc$ halos tend to point towards each other and the amplitude of this alignment is clearly enhanced for more massive halos.

To investigate the extent to which clusters move along the line of separation we determine the average velocity of halos along the connecting line as a function of separation. We determine the velocities of each halo along the connecting line by first calculating the direction of the line joining the two halos and then calculating the velocity component along this connecting line. These velocities are then binned logarithmically according to separation to obtain the mean streaming motion of the halos. The results are shown in Figure~\ref{fig:vel}. Halo pairs within the considered mass and separation ranges tend to move towards each other with increasing velocities as the distance between them decreases. This result compares well with results published by \cite{Sheth-Zehavi-08}.

\begin{table}
 \centering
 \caption{The velocity parallel to the orientation of the cluster divided by the velocity perpendicular to the cluster for the various mass ranges.}
 \begin{tabular}{lccc}
 \hline 
  & \multicolumn{3}{c}{Mass range($\hMsol$)}\\ 
 \cline{2-4}\\
  & $10^{12}-10^{13}$ & $10^{13}-10^{14}$ & $10^{14}-<10^{15}$ \\ 
 \hline 
 $\frac{V_\parallel}{V_\perp}$ & 1.051 & 1.072 & 1.100 \\ 
 \hline
 \end{tabular}
 \label{table:meanvel}
\end{table}

\begin{figure}
  \epsfig{file=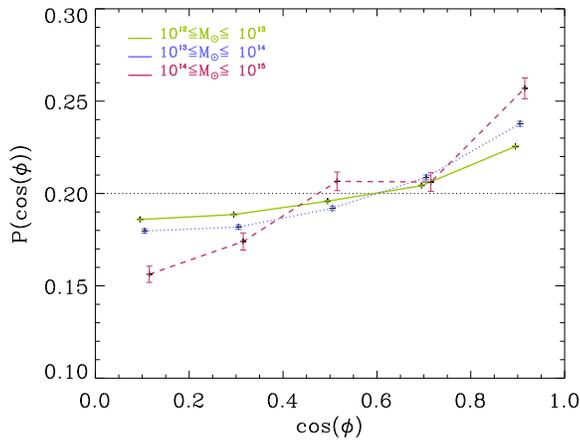,width=0.95\hsize}
  \caption{\label{fig:velax} Probability distribution for  the cosine of the angle, $\phi$, between the orientation and the velocity of halos within the indicated mass ranges. An isotropic distribution would result in the dotted horizontal line. Poissonian errors are shown.}
\end{figure}

Figure~\ref{fig:velax} displays the probability distribution of the cosine of $\phi$ where $\phi$ denotes the angle between the orientation of the halo and its bulk velocity. This was calculated by binning each of the halos according to its cosine of $\phi$. For all mass ranges we find an excess at large values of $\cos(\phi)$ indicating a preference for small angles between orientation and velocity. We see a correspondingly small number of halos with large angles between their orientations and velocities. A significant deviation from isotropy, displayed by the dotted horizontal line, for all mass ranges is apparent. Although this deviation increases with halo mass, in general the orientations of halos above $10^{12}\hMsol$ tend to be aligned with their velocities. We note that observations of the SZ effect are mostly only possible in clusters with masses $M\gtrsim10^{14}\hMsol$ where the effect is most pronounced. For this reason, in the following section where we investigate the kSZ effect, we do so only for the most massive halos.

The extent to which the velocities are aligned with the orientation has been quantified in Table~\ref{table:meanvel} where the average velocity parallel to the major axis for each mass bin is calculated and divided by the average velocity perpendicular to the major axis. One can see that for larger halos the velocity parallel to the major axis is on average 10\% higher than the velocity perpendicular. This trend is smaller for smaller mass halos as their elongation is less dramatic.

\section{Impact on kSZ measurements}
\label{sec:ksz}
The kSZ effect is proportional to the optical depth of a halo, $\tau$, and the velocity of the halo along the line of sight, $v_\parallel$.
The optical depth is defined by
$$\tau = \sigma_T \int n_e(l)\text{d}l \approx \frac{\sigma_T}{m_p} \frac{\Omega_b}{\Omega_0} \int \text{d}l \rho_{dm}(l)$$
where $\sigma_T$ is the Thomson scattering cross section, $n_e$ is the electron number density along the line of sight and $m_p$ is the mass of the proton. In defining this we assume, as outlined in \cite{Diaferio-Sunyaev-Nusser-00}, that the electron number density is related to the dark matter mass density along the line of sight, $\rho_{dm}(l)$, by the following equation
$$n_e = \frac{\rho_{dm}(l)}{m_p} \frac{\Omega_b}{\Omega_0}$$
where $\Omega_b$ is the baryonic density and $\Omega_0$ is the mean matter density in units of the critical density. We have thus assumed that the gas traces the dark matter halos and relates the properties of those halos to the properties of the gas. It is important to note that the fraction of gas at the cores of clusters may be lower than the cosmic value by $\sim30\%$ (\citealt{Kravtsov-05,Ettori-06}). As a result of this, the amplitude of the kSZ signal as calculated below could also be reduced by this percentage. 
However, the relative ratios between the dispersion parallel and perpendicular to the line of sight and between the real and random samples as used in our investigation will not be affected.

Peculiar velocities along the line of sight, $v_\parallel$, will produce kSZ temperature fluctuations. This kinematic effect is given by
$$k_{SZ} = \frac{\Delta T}{T} = \frac{\Omega_b}{\Omega_0} \frac{\sigma_T}{m_p} \int^L_0 \frac{v_\parallel(l)}{c} \rho_{dm}(l) $$
where $v_\parallel(l)$ is the bulk velocity along the line of sight at $l$ and $c$ is the speed of light.

\begin{figure*}
 \epsfig{file=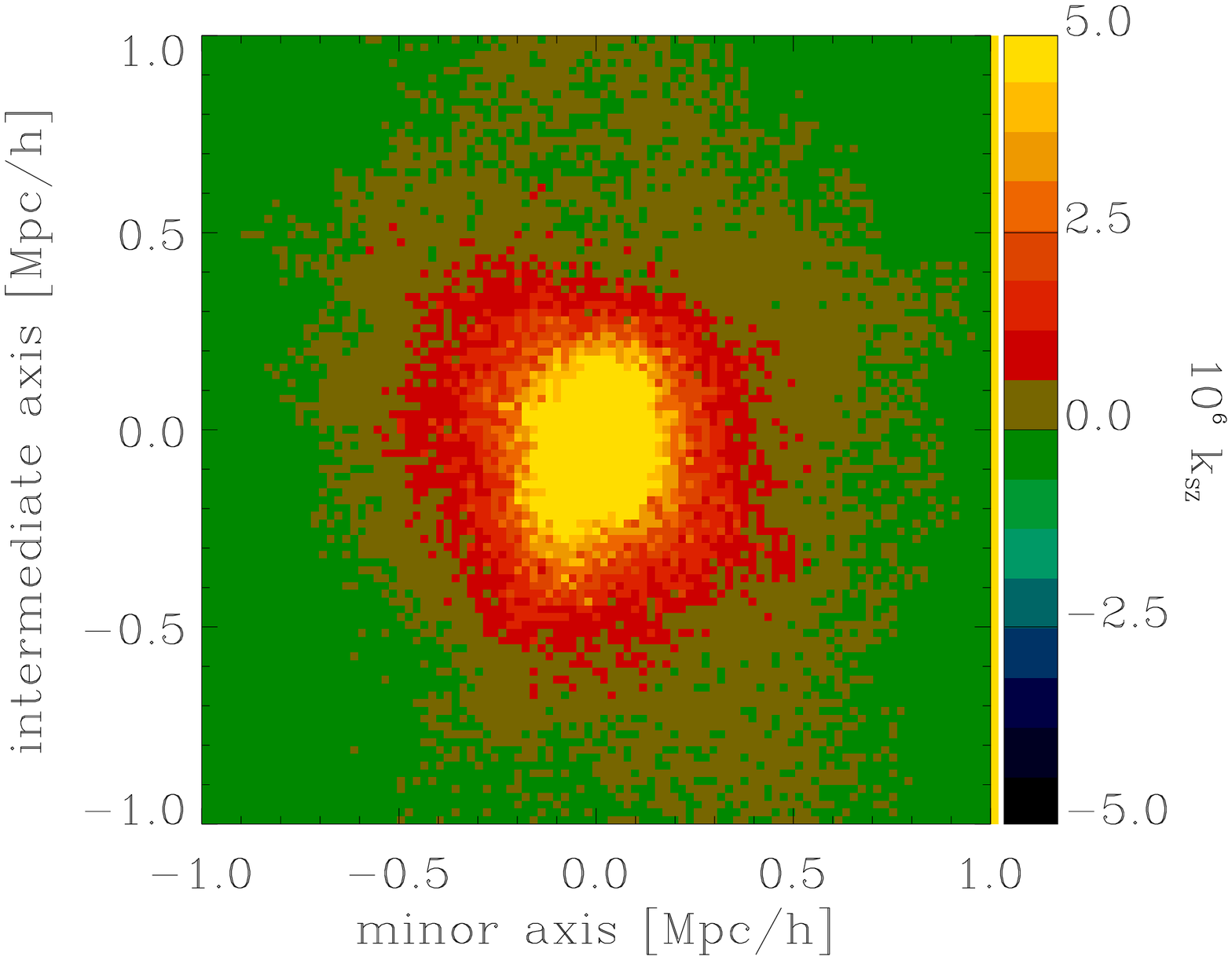,width=0.405\hsize}
\epsfig{file=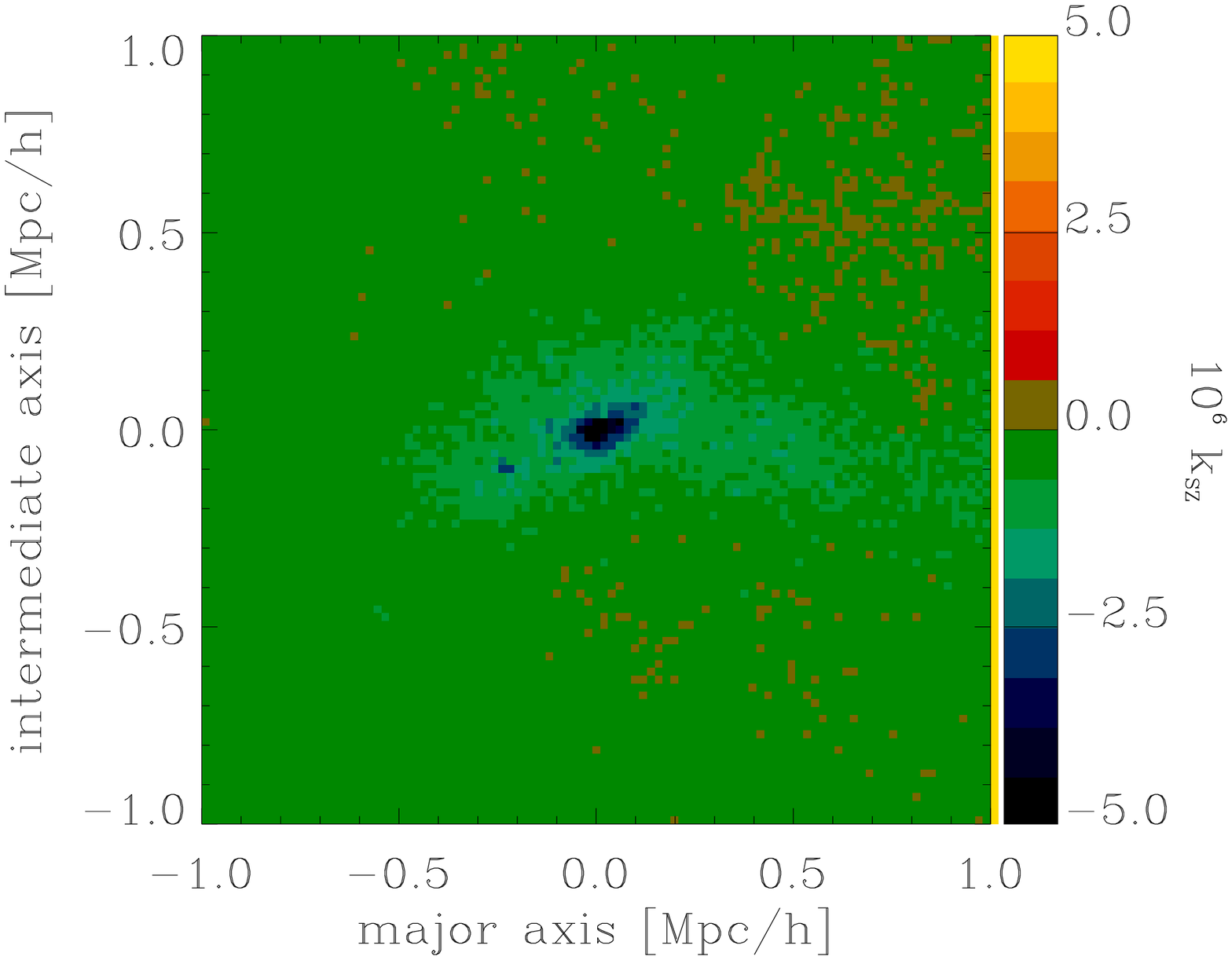,width=0.405\hsize}
\caption{\label{fig:example} Two projected kSZ maps of the same
halo  of   $1.2\times10^{14}\hMsol$.    Its  peculiar velocity  is
$640\kms$, the velocity  components along the major,  intermediate and
minor axes are $597\kms$, $57\kms$  and $-224\kms$, respectively. {\it
The left panel} shows a kSZ map  if the halo is  rotated such that the
major axis is  parallel to the line  of sight.  {\it The right  panel}
displays the map for the same halo with  the minor axis oriented along
the line of sight. For illustrative purposes we here use much higher
resolution than that used for the determination of $\sigma_{kSZ}$ shown in Figure \ref{fig:ksig}. }
\end{figure*}

We calculate $k_{SZ}$ for halos in our simulation by placing a box of size $5.5\hMpc$ 
centered on each halo and dividing it into cells of various lengths. The spatial resolution of the simulation is $5\hkpc$ \citep{Springel-05a} while the best angular resolution possible with current telescopes ($\sim1$ arcminute) corresponds to about $40\hkpc$ in the nearest clusters. Therefore we have investigated a range of pixel sizes from $50\hkpc$ upwards. The box size and the number of cells per dimension were chosen to ensure that the center of halo coincides with the center of the most central cell. The integral is calculated by summing the contributions from discrete cells along one axis which corresponds to the line of sight, $l$. Hence we obtain 2D maps of the kSZ signal for each halo. Figure~\ref{fig:example} shows an example of the kSZ signal of a cluster-sized halo from two viewpoints. The maximum value of the kSZ signal in the map, which in all cases comes from the central pixel, is taken as the amplitude of the kSZ signal for that halo and, to relate this to work on the power spectrum of kSZ fluctuations, we calculate $\sigma_{kSZ}$, the root mean square fluctuation in this amplitude. This approach is similar in nature to the peak model presented by \cite{Yoshida-Sheth-Diaferio-01}. Using their arguments we are thus confident that possible contributions resulting from secondary structures such as filaments may be neglected.

The kSZ signature of a cluster elongated along the line of sight is enhanced relative to a spherical cluster of the same mass because it has a larger optical depth. To separate this effect from the additional enhancement due to the velocity correlation, we calculated the kSZ velocity dispersion firstly for the correct halo peculiar velocities, and secondly for the case where the directions of the velocities are randomized. The transformation from real to random bulk velocities has to be performed on a particle by particle basis. First, the bulk velocity is subtracted from each particle's velocity and after a random rotation it is added to each of the particles again. These new velocities are used in the integration along the line of sight. The kSZ signal is presented in units of $[\mu K/K]$ for comparison with previous work \citep{Yoshida-Sheth-Diaferio-01}.

\begin{figure}
  \epsfig{file=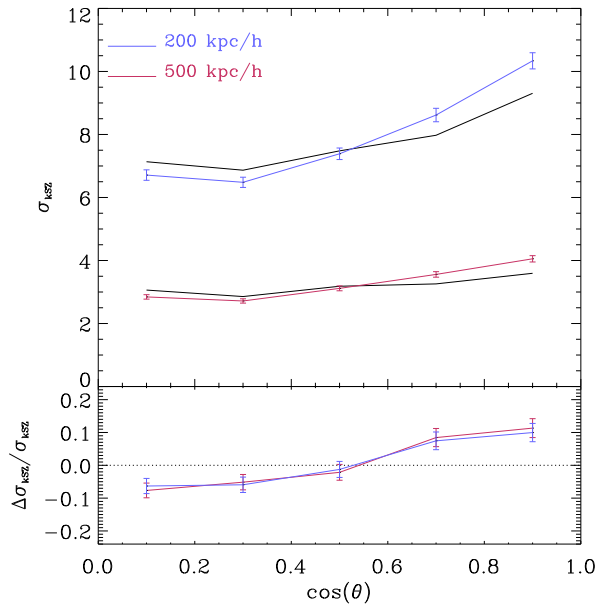,width=0.95\hsize}
  \caption{\label{fig:ksig} {\it Upper panel:} the rms fluctuations on a $200\hkpc$ and $500\hkpc$ scale in the kSZ signal as a function of the cosine of the angle between halo orientation and the line of sight. Results are plotted for only the largest mass range ($M\gtrsim 10^{14} \hMsol$). The coloured lines show the $\sigma_{kSZ}$ obtained using the velocities measured in the simulation. The black lines show the $\sigma_{kSZ}$ obtained when the bulk velocities are randomized. {\it Lower panel:} the difference between the measured kSZ signal and the randomized signal (normalized by the randomized signal), illustrating the effect of the velocity-shape alignment}
\end{figure}

The results for two different pixel sizes are illustrated in Figure~\ref{fig:ksig} where the coloured lines show the kSZ velocity dispersion for the true velocities of the halos, as a function of the angle between halo orientation and the line of sight. The peak signal is enhanced by up to 60\% for halos where the orientation is along the line of sight compared to those oriented perpendicular to the line of sight. 
The black line in Figure~\ref{fig:ksig} represents the kSZ signal when the cluster velocities are randomized. We observe that randomizing the velocities diminishes the peak signal by approximately 20\%. The observed behavior indicates that the dependence of the kSZ signal dispersion on the orientation of the clusters relative to the line of sight is due to both the change in optical depth (if the cluster is resolved) and the velocity--shape alignment.
The lower panel of Figure~\ref{fig:ksig} displays the residual $\Delta \sigma_{\rm kSZ}/\sigma_{\rm kSZ}$. 

To see the effect of resolution we have computed the kSZ signal for additional cell/pixel sizes, including $50$, $1000$ and $2000\hkpc$ on a side. For $L=50\hkpc$, we find higher peaks in the kSZ signal and an enhanced dispersion due to the larger optical depth for the central pixels. 
However, the dependence on orientation relative to the line of sight and the relative difference to the cluster sample with randomized velocities does not change. If the cell size used in the integration along the line of sight is large enough, we expect to include most of the dark matter halo and the velocities associated with it, thereby excluding effects due to advantageous orientation. We investigated the effect for pixel sizes of $L=1$ and $2\hMpc$ and confirmed that although the amplitude of the kSZ signal was reduced and the dependence on orientation was slightly weakened the relative difference to the randomized velocity sample, $\Delta \sigma/\sigma$, did not change significantly. For $L = 2\hMpc$ pixel sizes we observed only 20\% difference between the kSZ dispersions for clusters oriented parallel and perpendicular to the line of sight. For the corresponding randomized velocity sample the kSZ dispersion is largely independent of orientation. Since at that resolution the entire cluster is covered by one pixel and the optical depth is independent of cluster's orientation, we can be confident that the observed dependence of the kSZ dispersion on the orientation relative to the line of sight is due only to the velocity--shape alignment.
\section{Conclusion}
\label{sec:conclusion}
Clusters of galaxies have been shown to both point towards each other and have a tendency to stream towards each other. Motivated by this we have investigated the correlation between the orientation of a cluster and its mean peculiar velocity. We have found in general that this correlation is significant. For massive clusters we have observed that the velocity parallel to the major axis is on average 10\% higher than velocity perpendicular to the orientation of the cluster (Table~\ref{table:meanvel}).

We have then explored how the orientation of clusters effects the kSZ signal and found that there is a 60\% enhancement of the kSZ signal dispersion for massive clusters which are orientated along our line of sight compared to those orientated perpendicular to the line of sight. This value is based on a pixel size of $L=200\hkpc$. The difference for unresolved clusters, $L=2\hMpc$, is 20\%. A smaller, but significant, effect for less massive clusters can also be seen.  

The enhancement of the kSZ signal is a result of the alignment of both the orientation and the velocity with the line of sight. In order to disentangle these effects and quantify the effect of the velocity alone we introduced a random sample. In this sample the orientations of the clusters remained fixed and their bulk velocities were randomized. We observe a 20\% difference in the kSZ dispersion for the random sample which is caused by the dependence of the optical depth of the cluster relative to the line of sight alone, indicating that the remaining 20\% of the enhancement is due to the velocity-shape alignment.

We have observed a correlation between the orientation and the velocities of clusters on scales up to $100\hMpc$. Such large-scale correlations, particularly in small-scale surveys, may result in a large bias in the kSZ signal depending on the orientation of large scale structures relative to the line of sight. In addition, since we have shown up to a 60\% increase in the kSZ signal in a line-of-sight cluster, the selection of kSZ sources may be substantially biased towards clusters aligned along the line of sight. 

\section*{Acknowledgments}
We thank the South African Centre for High Performance Computing, National Institute of Theoretical Physics, Square Kilometer Array Project and National Research Foundation for support. 
\bibliography{lit}
\label{lastpage}

\end{document}

%% file: macro.tex
\newcommand{\kms}{\>{\rm km}\,{\rm s}^{-1}}

\newcommand{\hMsol}{{\>h^{-1}\rm M}_\odot}
\newcommand{\hMpc}{{\>h^{-1}\rm Mpc}}
\newcommand{\hkpc}{{\>h^{-1}\rm kpc}}

\newcommand{\beq}{\begin{equation}}
\newcommand{\eeq}{\end{equation}}

\newdimen\hssize
\newdimen\hdsize 